\documentclass{article}

\usepackage{arxiv}

\usepackage[utf8]{inputenc} % allow utf-8 input
\usepackage[T1]{fontenc}    % use 8-bit T1 fonts
\usepackage{hyperref}       % hyperlinks
\usepackage{url}            % simple URL typesetting
\usepackage{booktabs}       % professional-quality tables
\usepackage{amsfonts}       % blackboard math symbols
\usepackage{nicefrac}       % compact symbols for 1/2, etc.
\usepackage{microtype}      % microtypography
\usepackage{lipsum}		% Can be removed after putting your text content
\usepackage{graphicx}
\usepackage{natbib}
\usepackage{doi}

\title{Optimality in superselective surface binding by multivalent DNA nanostars}

\author{ \href{}{\includegraphics[scale=0.06]{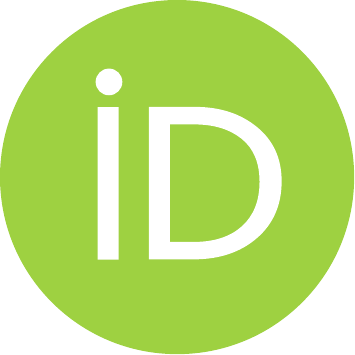}\hspace{1mm}Christine ~Linne} \\
	Department of Bionanoscience\\ 
    TU Delft, 2629 HZ Delft, The Netherlands\\
    Soft Matter Physics, Huygens-Kamerlingh Onnes Laboratory\\
    Leiden Institute of Physics,2300 RA Leiden, The Netherlands\\
    \texttt{C.Linne@tudelft.nl}
     \\
	\And
	\href{}{\includegraphics[scale=0.06]{orcid.pdf}\hspace{1mm}Eva ~Heemskerk} \\
	Department of Bionanoscience\\ 
    TU Delft, 2629 HZ Delft, The Netherlands\\
    \texttt{E.M.Heemskerk@student.tudelft.nl}
 \And
	\href{}{\includegraphics[scale=0.06]{orcid.pdf}\hspace{1mm}Jos ~Zwanikken} \\
	Department of Bionanoscience\\ 
    TU Delft, 2629 HZ Delft, The Netherlands\\
    \texttt{J.Zwanikken@tudelft.nl}
 \And
	\href{}{\includegraphics[scale=0.06]{orcid.pdf}\hspace{1mm} Daniela J. ~Kraft} \\
	Soft Matter Physics, Huygens-Kamerlingh Onnes Laboratory\\
    Leiden Institute of Physics,2300 RA Leiden, The Netherlands\\
	\texttt{Kraft@physics.leidenuniv.nl} \\
 \And
	\And
	\href{}{\includegraphics[scale=0.06]{orcid.pdf}\hspace{1mm}Liedewij ~Laan} \\
	Department of Bionanoscience\\ 
    TU Delft, 2629 HZ Delft, The Netherlands\\
	\texttt{L.Laan@tudelft.nl} \\
}

\renewcommand{\shorttitle}{superselectivity by DNA nanostars}

\hypersetup{
pdftitle={A template for the arxiv style},
pdfsubject={q-bio.NC, q-bio.QM},
pdfauthor={L ~Laan},
pdfkeywords={First keyword, Second keyword, More},
}

\begin{document}

\maketitle
\begin{abstract}
	Weak multivalent interactions govern a large variety of biological processes like cell-cell adhesion and virus-host interactions. These systems distinguish sharply between surfaces based on receptor density, known as superselectivity. Earlier experimental and theoretical work provided insights into the control of selectivity: Weak interactions and a high number of ligands facilitate superselectivity. Present experimental studies typically involve tens or hundreds of interactions, resulting in a high entropic contribution leading to high selectivities. However, if, and if so how, systems with few ligands, such as multi-domain proteins and virus binding to a membrane, show superselective behavior is an open question.  Here, we address this question with a multivalent experimental model system based on star shaped branched DNA nanostructures (DNA nanostars) with each branch featuring a single stranded overhang that binds to complementary receptors on a target surface. Each DNA nanostar possesses a fluorophore, to directly visualize DNA nanostar surface adsorption by total internal reflection fluorescence microscopy (TIRFM). We observe that DNA nanostars can bind superselectively to surfaces and bind optimally at a valency of three. We quantitatively explain this optimum by extending the current theory with interactions between DNA nanostar binding sites (ligands). Our results add to the understanding of multivalent interactions, by identifying microscopic mechanisms that lead to optimal selectivity, and providing quantitative values for the relevant parameters. These findings inspire additional design rules which improve future work on selective targeting in directed drug delivery.

\end{abstract}

% keywords can be removed

\section{Introduction}
\begin{figure}
    \centering
    \includegraphics[width=100mm]{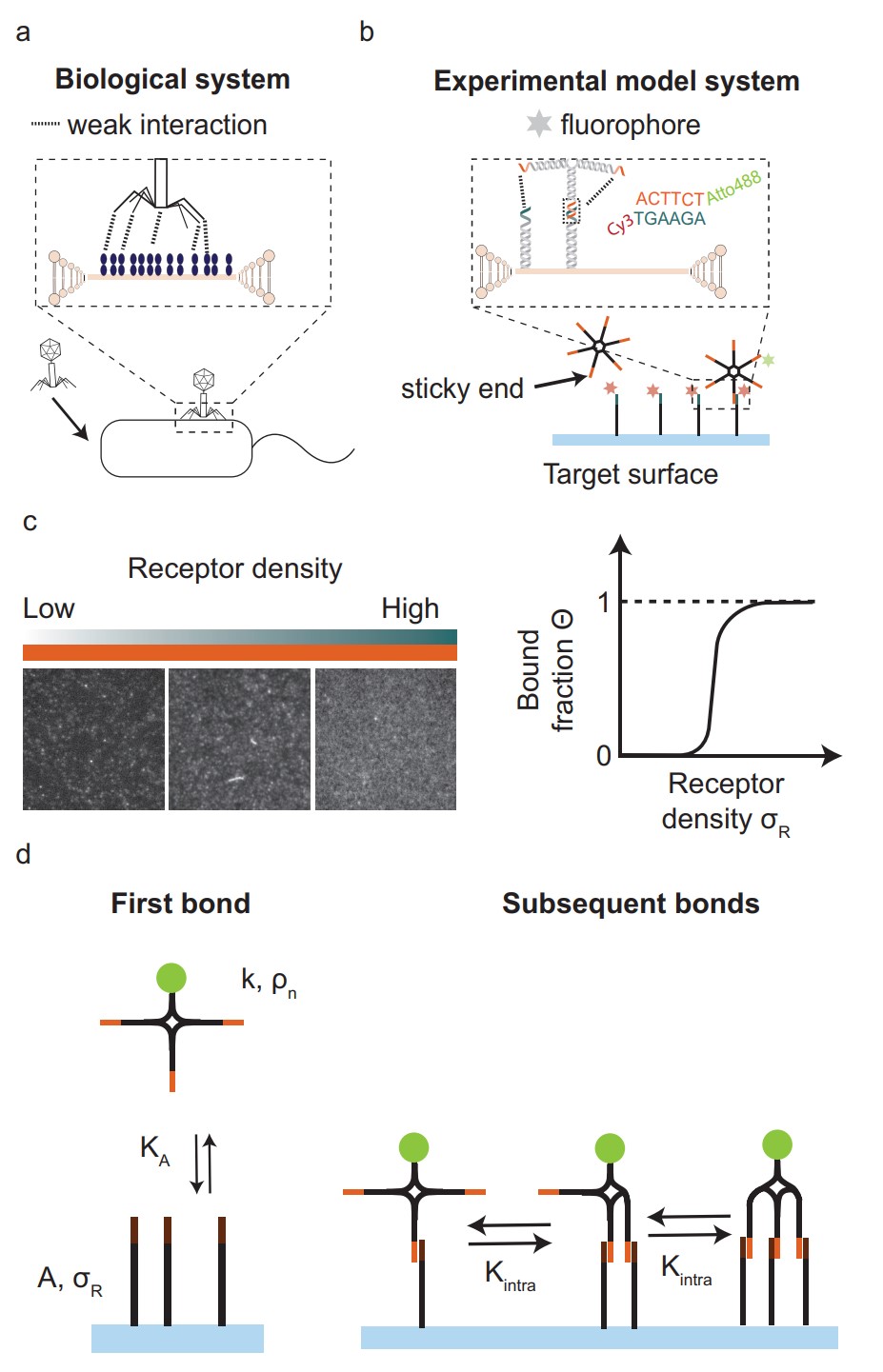}
    \caption{\textbf{Motivation and model system} a) virus-cell adhesion as an example where objects with <10 interaction sites participate in superselective surface binding. b) DNA nanostars as an experimental model system for valencies below $10$, that can bind to a supported lipid bilayer (SLB). Each arm of the DNA nanostar features a single stranded overhang (sticky end) that binds to the complementary sticky end on the receptors in the SLB. Each DNA nanostar possesses a fluorophore, such as Cy3 or Atto488, attached to one arm. c) TIRFM images that show how the change in the number of adsorbed DNA nanostars on the target surface is quantified for variable number of receptors on the surface. After background subtraction and normalisation with the saturated value, we translate the signal into a bound fraction $\Theta$ and plot it against the receptor density $\sigma_{\rm R}$ to determine the selectivity $\alpha$. d) cartoon of the theory used to describe multivalent surface binding by DNA nanostars.}
    \label{fig:Intro_bio_model_system}
\end{figure}

{M}ultivalent interactions, where multiple ligands and receptors together form a single bond, are ubiquitous in nature. For example during bond formation by intrinsically disordered protein-protein interactions \cite{Teilum2021,Volkov2018, Maan2022, Banjade2014PhaseReceptors}, ubiquitylation \cite{Liu2010}, antibody-antigen binding \cite{Goldberg2002} and virus-host binding \cite{Overeem2021, Overeem2020}(Fig.~\ref{fig:Intro_bio_model_system}a). In these examples individual ligand-receptor interactions are weak and highly reversible but together they establish a strong and often highly specific bond. 

To understand how biological systems achieve high selectivity upon binding, Martinez-Veracoechea and Frenkel introduced the concept of superselectivity as a non-linear increase in the binding probability \cite{Martinez-Veracoechea2011}. In their model a multivalent particle distinguishes surfaces based on receptor density. A change in the interaction strength, valency and/or particle concentration manipulates the sharpness of this transition, where high valency, weak interactions and low particle concentrations yield the highest selectivity. In addition, recent studies illustrated that parameters like crowding  \cite{Christy2021}, the addition of an external force on the particle \cite{Curk2020} and competition \cite{Curk2022} also regulate selectivity.

Present experimental systems that successfully demonstrated multivalent surface binding include polymers \cite{Dubacheva2015,Dubacheva2014}, viruses \cite{Overeem2021} and nano- \cite{Lanfranco2019, Phan2023} and microparticles \cite{Linne2021,Scheepers2020}. These systems either feature hundreds of interaction sites, as is the case for polymers and nano- and colloidal particles, or have limited experimental control over interaction strength and valency, as is the case for virus particles that interact with less than $10$ receptors \cite{Szklarczyk2013}. 

In this paper we focus on multivalent surface binding by systems with few interaction sites, which occur, for example, during virus host binding \cite{Overeem2021}, during binding of microtubules to chromosomes in mitosis \cite{Volkov2018} or when multi-domain proteins bind to the cell membrane during polarity establishment in development\cite{lang_oligomerization_2022}. In addition, systems with few ligands can provide insights in the transition from monovalent to multivalent binding. Specifically we ask how selectivity of binding to  receptor-covered surfaces by multivalent systems with few ligands depends on valency, interaction strength and physical properties of the system. 

To experimentally address this question precise control is needed of valency, ligand and receptor interaction strength and the particle's concentration. In addition, control over other physical properties of this multivalent system, such as the flexibility of the ligands, self-interactions, and pair-interactions between the ligands is desirable for optimising the conditions for superselectivity. Here we exploit an experimental model system of DNA-origami nanostars to experimentally assess superselectivity in systems with low valencies and with tunable binding strength. A DNA nanostar consists of branched junctions of DNA strands also called arms with single stranded sticky overhangs that act as binding sites (ligands) \cite{Conrad2019,Brady2019,Biffi2013PhaseNanostars}. The sticky ends on each DNA nanostar bind to surface mobile complementary DNA strands (receptors), see Fig.~\ref{fig:Intro_bio_model_system}b. DNA nanostars have a number of attractive features: the length of the sticky end regulates the interaction strength, the number of arms precisely dictates the valency and a fluorophore attached to one arm allows for the visualization of DNA nanostar-surface adsorption with Total Internal Reflection Fluorescence Microscopy (TIRFM).

We observe that multivalent DNA nanostars can bind superselectively to a surface coated with laterally mobile receptors, and we find that both valency and binding strength have an optimum for superselective surface binding. We extend the current theory by including interactions between DNA nanostars arms to be able to quantitatively match the observations, and find that the ligand pair-interaction strength has an optimum value as well for achieving maximal selectivity. From the extended model, we derive additional design rules for superselective surface binding and discuss what our findings could imply for biological systems.

% how strong are the bonds and what determines the strength?
% how can a system both be superselective and strongly bound at the same time
% naively: number of bonds

\section{Results}
\label{sec:nanostar_noinert}

%\begin{figure}[htbp]
%    \centering
%    \includegraphics[width=0.8\textwidth]{Figures/Results/20211130_theory.eps}
%    \caption{\textbf{Model for $k=6$ and different $K_{\rm A}$ and $K_{\rm intra}$.} a) Theoretical prediction of $\Theta$ for a monovalent particle for different $K_{\rm A}$, see Eq.~\ref{eq:theta_mono}.b) Eq.~\ref{eq:theta_multi} predicts $\Theta$ for fixed $K_{\rm A} = 1e-05~\rm M^{-1}$ and varying $K_{\rm intra}$. c) Average number of bound arms $\langle n \rangle$ for varying $K_{\rm intra}$, see Eq.~\ref{eq:arms}.}
%    \label{fig:theory}
%\end{figure}

%In this paper we present how varying the valency and interaction strength affects surface binding using a multivalent DNA nanostar system. We will fit and interpret our experimental results in the context of the model by Frenkel and coworkers, which we first introduce, followed by the description of our model system and our results. Based on our experimental data we extend the model and explore the implications of these exptensions with computer simulations.

\section*{DNA nanostars as an experimental model system}

Our experimental system to systematically study superselective surface targeting with 1-10 ligands consists of DNA nanostars. To quantitatively elucidate the transition of DNA nanostar adsorption, we employed DNA nanostars with different number of arms $k$, and imaged their adsorption to supported lipid bilayers (SLBs) functionalized with different receptor concentrations $\sigma_{\rm R}$. We employ DNA nanostars with a ssDNA sequence at the end of each arm (sticky end) that binds to receptors on a target surface with the complementary sticky end, see Fig.~\ref{fig:Intro_bio_model_system}b. The receptors consist of a $77~$bp double stranded stem with a cholesterol molecule on the $5'$ end. Cholesterol integrates the receptor into a SLB on the target surface and ensures full mobility of the receptors \cite{VanDerMeulen2014,Rinaldin2019ColloidSelf-assembly,Chakraborty2017ColloidalFlexibility}. On the $3'$ end the receptors have a sticky end with the complementary sequence to the DNA nanostar sticky end. The length of the sticky end determines the hybridization free energy of each arm, see Materials and Methods for details. Each DNA nanostar possesses an Atto488dye on the $3'$ end of one arm, which does not inhibit binding. The excitation of the fluorophore and acquisition of the emission with total internal reflection microscopy (TIRFM) allows for the direct visualization of DNA nanostar-surface adsorption. The advantage of TIRFM is the direct excitation of the DNA nanostars on the surface and limited excitation of DNA nanostars in solution. We imaged the DNA nanostar signal for different receptor densities ranging from low to high, see (Fig.~\ref{fig:Intro_bio_model_system}b). We measured the mean intensity of a certain area size, and normalised it with respect to the maximum intensity of the same area size. Finally, we plot the normalised signal, which is equal to the bound fraction $\Theta$ as a function of $\sigma_{\rm R}$, see (Fig.~\ref{fig:Intro_bio_model_system}b). 

 \section*{Optimal valency for superselective surface binding}
\begin{figure}
    \centering
    \includegraphics[width=100mm]{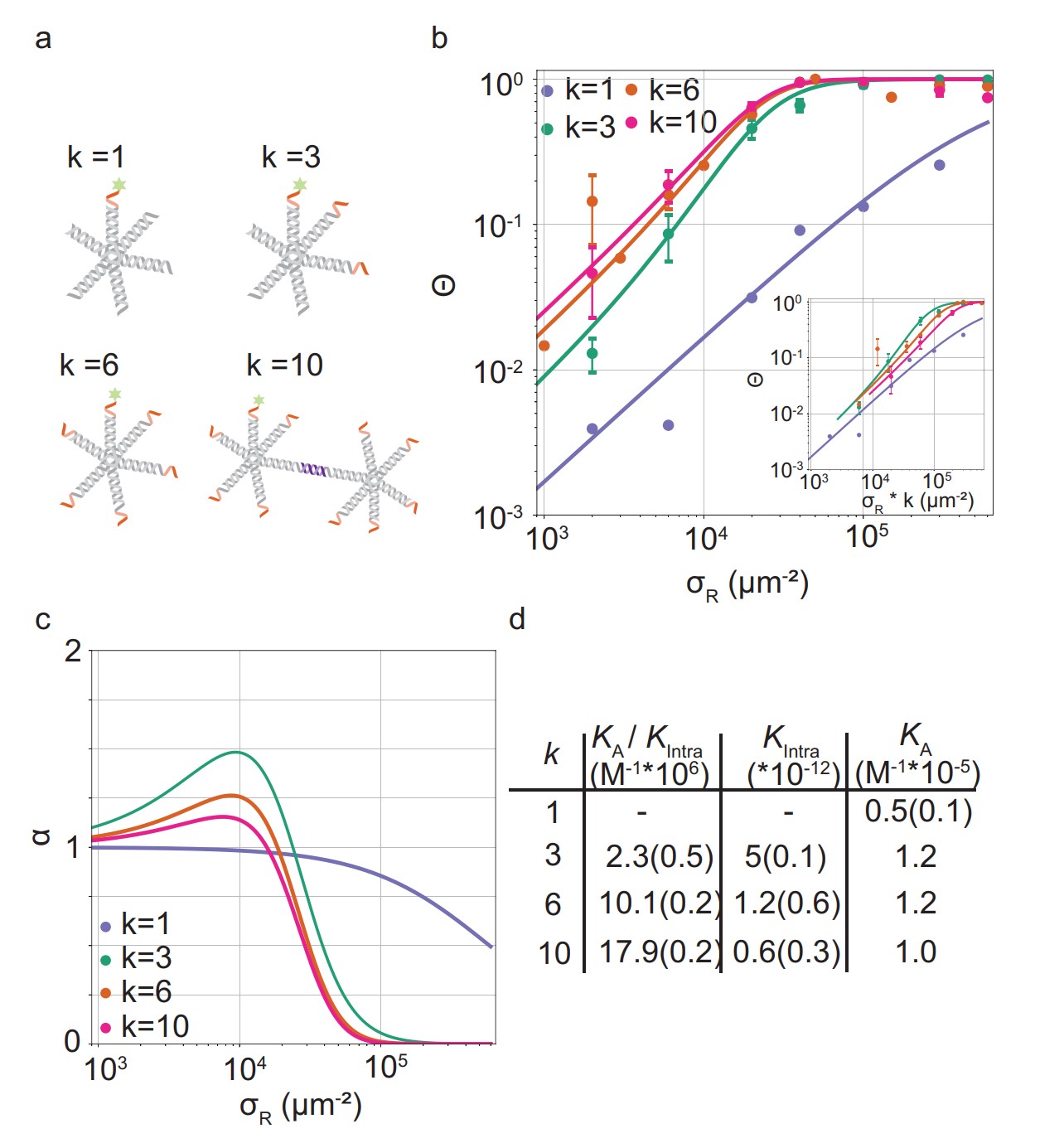}
    \caption{\textbf{Optimality in valency of DNA nanostar surface binding} a) Cartoons of DNA nanostars with a valency $k$ of 1, 3, 6 or 10. The red parts indicate the single stranded binding sites. The fluorophore is depicted in green. b) The bound fraction $\Theta$ measured as a function of receptor density $\sigma_{\rm R}$ for sticky end ACTTCT and four valencies $k=1,3,6,10$. the lines are least-squared fits of the model Eq.~\ref{eq:theta_multi} adapted from Frenkel and coworkers \cite{Martinez-Veracoechea2011} with fitting parameters $K_{\rm A}$ and $K_{\rm intra}$. For $k=3$ we excluded the last four datapoints for the fit to get the most accurate mathematical description of the non-linear transition of $\Theta$. The insert shows the bound fraction $\Theta$ measured as a function of the receptor density $\sigma_{\rm R}$ rescaled with $k$.
    c) The selectivity parameter $\alpha = \frac{\rm d \ln\Theta}{\rm d \ln \sigma_{\rm R}}$ $\alpha$ as a function of receptor density $\sigma_{\rm R}$, shows an optimal  $\alpha$ for $k=3$. d) Table with the bindings constants obtained from a fit with Eq.~\ref{eq:theta_multi} to $\Theta$ in Fig.~\ref{fig:superselectivity_noinert}b yields the fitting parameters $K_{\rm A}/K_{\rm intra}$ and $K_{\rm intra}$. The numbers in brackets indicate the fitting error. A division of the two fitting parameters yields $K_{\rm A}$.}
    \label{fig:superselectivity_noinert}
\end{figure}

We started by investigating the DNA nanostar-surface adsorption for different number of arms $k = 1,3,4,10$ but with fixed hybridization energy $\Delta G^0 = -7~\rm{k_BT}$ at a constant DNA nanostar concentration in solution $\rho_n = 10^{-8}~$M (Fig.~\ref{fig:superselectivity_noinert}a). We choose this interaction strength based on our findings in a previous study with colloidal particles \cite{Linne2021}, where superselective binding occurred for ligand-receptor interaction strengths around $\Delta G^0 = -7~\rm{k_BT}$. We measured $\Theta$ over receptor densities ranging between $\sigma_{\rm R} = 1.000 - 600.000~\rm {\mu m^{-2}}$, see Fig.~\ref{fig:superselectivity_noinert}b, and found that $\Theta$ smoothly increases with increasing $\sigma_{\rm R}$. With increasing valency $k$ the curve shifts to lower $\sigma_{\rm R}$. This can be understood because DNA nanostars with a valency of $k$ have a $k$ times higher ligand concentration at the same concentration as a monovalent DNA nanostar, which therefore implies that the binding probability of the DNA nanostar scales with $k$. To test this explanation for our experimental system, we multiply $\sigma_{\rm R}$ with $k$, see insert in Fig.~\ref{fig:superselectivity_noinert}b. From this plot we find that the multivalent binding curves shift towards the monovalent curve and fall on top of each other in the low  $\sigma_{\rm R}$ range. In this range, DNA nanostars most likely bind one arm only and thus effectively bind monovalently. However, at increasing $\sigma_{\rm R}$ the curves start to deviate from each other underlining that this non-linear increase should be caused by multivalent interactions. 

Next we determine how valency affects the selectivity of surface binding of DNA nanostars by fitting our experimental data with eq 2 and 5. Fig.~\ref{fig:superselectivity_noinert}c presents $\alpha$ for $\Theta$ in Fig.~\ref{fig:superselectivity_noinert}b. As expected for a monovalent DNA nanostar the selectivity never exceeds $1$, because it follows the Langmuir isotherm, which has a maximum slope of $1$. An increase in $k$ is expected to lead to an increase in $\alpha$. In Fig.~\ref{fig:superselectivity_noinert}c we indeed observe superselective behavior ($\alpha$ > 1) for our multivalent DNA nanostars and we observe an optimum in selectivity for $k=3$. Interestingly, previous computational work on systems with many binding sites also showed an optimum in valency \cite{Martinez-Veracoechea2011}.

To investigate the origin of this optimum, we also determine the chemical equilibrium constants $K_{\rm A}$ and $K_{\rm intra}$ using the theory previously developed by Frenkel et al \cite{Martinez-Veracoechea2011}.  
Their simplest model describes the adsoption of monovalent DNA nanostars as a Langmuir isotherm \cite{Martinez-Veracoechea2011, Curk2018} which is written in the specific form
$$ \Theta = \frac{ \rho_n A \sigma_{\rm R} K_{\rm A}}{1 + \rho_n A \sigma_{\rm R} K_{\rm A}}
    \label{eq:theta_mono}$$

where $\Theta$  is the bound fraction, $\rho_n$ is the DNA nanostar concentration in solution, $\sigma_{\rm R}$ is the receptor density on the target surface, $A$ is a unit surface area and $K_{\rm A}$ is the equilibrium association constant to form a single bond, see (Fig.~\ref{fig:Intro_bio_model_system}c). The equilibrium constant $K_{\rm A}$ determines the specific concentration of receptors where half of the DNA nanostars are bound. In the expression we exchanged the activity $z$ of the DNA nanostars by the concentration $\rho_n$, given that the low concentrations in the experiments are in the nM-range.

The interaction strength between ligands and receptors shifts  $\Theta$ relative to $\sigma_{\rm R}$, and larger interaction strengths (effectively described by $K_{\rm A}$) shift the transition point to lower concentrations. The number of arms $k > 1$ introduces a combinatorial term to the system that accounts for the number of possible bond formations, see (Fig.~\ref{fig:Intro_bio_model_system}c). The extra degrees of freedom of multivalent DNA nanostars add to specific entropy and energy differences between the bound states. Assuming that the arms act independently, so ignoring any (effective) attractions or repulsions between them, the adsorption can then be written as
$$
    \Theta = \frac{\rho_n K^{\rm av}_{\rm A} (K_{\rm A}, K_{\rm intra}, k)}{1 + \rho_n  K^{\rm av}_{\rm A}(K_{\rm A}, K_{\rm intra}, k)},
    \label{eq:theta_multi}
$$
with $K^{\rm av}_{\rm A}$, the equilibrium avidity association constant, and $K_{\rm intra}$ is related to the equilibrium constant for the second bond, after the first bond is established (this constant would be $(k-1) K_{\rm intra}/2$). The constant $K^{\rm av}_{\rm A}$ represents the formation rate of the first bond and, additionally, includes a combinatorial term that describes the formation of subsequent bonds:
$$
    K^{\rm av}_{\rm A} = \frac{K_{\rm A}}{K_{\rm intra}} \left[ (1+\sigma_{\rm R} A K_{\rm intra})^k - 1\right].
$$

As described above, if the bonds are formed independently of each other, and if the DNA nanostars effectively behave like monovalent particles, the binding probability of the multivalent particles scale with the valency $k$:
$$
    \Theta_{\rm multi} = k * \Theta_{\rm mono}.
    \label{eq:k_shift}
$$

The selectivity $\alpha$ of a multivalent system quantifies how sharp $\Theta$ increases with receptor density $\sigma_{\rm{R}}$ \cite{Martinez-Veracoechea2011}: 
$$
    \alpha  = \frac{{\rm d\ln}\Theta}{{\rm d\ln}\sigma_{\rm R}}.
    \label{eq:alpha}
$$
More precisely, $\alpha$ describes the slope of $\Theta$ as a function of $\sigma_{\rm R}$ on a log-log scale and $\alpha > 1$ defines a superselective system. One can also interpret $\alpha$ as the (density-dependent) `Hill-coefficient', with $\Theta \propto \sigma_{\rm{R}}^\alpha$.  with a least-square fit of Eq.~\ref{eq:theta_multi} to the experimental data in Fig.~\ref{fig:superselectivity_noinert}b. The fitting results are reported in Tab.~\ref{tab:Fitting_par_Fig_3}d.

The comparison of $K_{\rm A}$ for $k=1,3,6,10$ in Tab.~\ref{tab:Fitting_par_Fig_3}d reveals no significant difference, consistent with the assumption we used in our model. Subsequent bond formations are captured by the second fitting parameter $1/K_{\rm intra}$. Interestingly, we observe a decrease in $K_{\rm intra}$ with decreasing $k$. This suggests that DNA nanostars with more arms are less likely to bind with multiple arms to the SLB, which we will follow-up on later in this paper.

\section*{Optimal binding strength for superselective surface binding}

%\begin{figure}[htbp]
%    \centering
%    \includegraphics[width=0.8\textwidth]{Figures/Results/20211119_theory_weak_strong.eps}
%    \caption{\textbf{Predicting $\Theta$ for strong and weak binding.} Comparing the adsorption curves for DNA nanostars with the same geometry ($k=6$) for weak (a) and strong (b) binding for increasing $K_{\rm intra}$.}
%    \label{fig:theory_weak_strong}
%\end{figure}

As a next step we studied the impact of interaction strength on superselectivity and the equilibrium binding constants. The model predicts that weakening the interaction strength will shift the curve of $\Theta$ towards larger $\sigma_{\rm R}$. The region where $\Theta \ll 1$ needs to be sufficiently large to facilitate a non-linear transition, which is determined by $K_{\rm intra}$. 

\begin{figure}
    \centering
    \includegraphics[width=100mm]{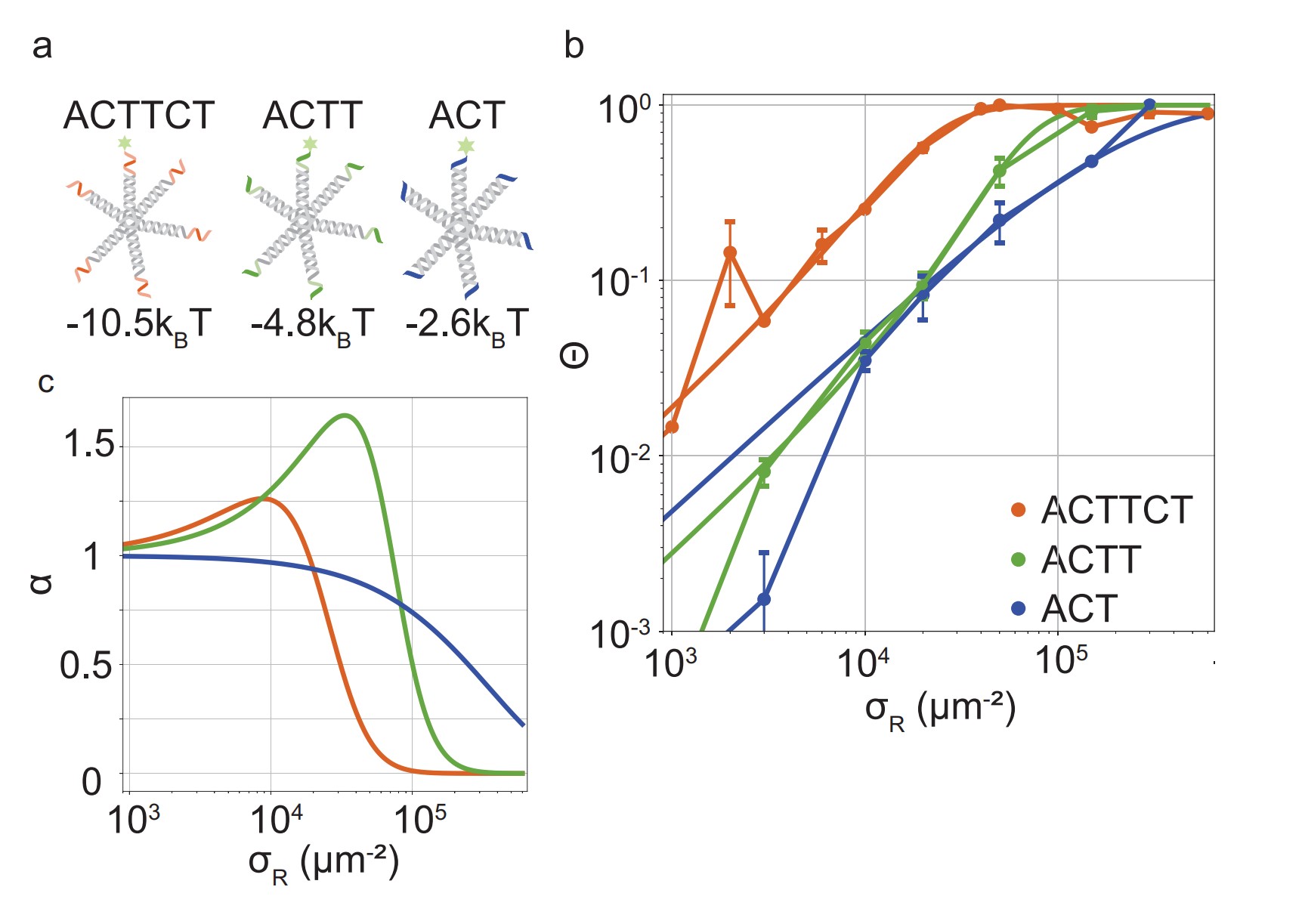}
    \caption{\textbf{Optimality in binding strength of DNA nanostar surface binding} a) Cartoons of DNA nanostars with varying binding strength. b) The bound fraction $\Theta$ measured as a function of receptor density $\sigma_{\rm R}$ for sticky end ACTTCT and three binding strength. the lines are least-squared fits of the model Eq.~\ref{eq:theta_multi} adapted from Frenkel and coworkers \cite{Martinez-Veracoechea2011} with fitting parameters $K_{\rm A}$ and $K_{\rm intra}$. 
    c) The selectivity parameter $\alpha = \frac{\rm d \ln\Theta}{\rm d \ln \sigma_{\rm R}}$ $\alpha$ as a function of receptor density $\sigma_{\rm R}$, shows an optimal  $\alpha$ for a binding strength of $-4.8k_BT$.}
    \label{fig:6bp_vs_4bp}
\end{figure}

\begin{table}[]
    \centering
    \begin{tabular}{c|c|c|c|c}
         sticky end (bp) & $k$ & $K_A \ K_{intra} $ & $ K_{intra}$ & $K_A$ \\
         & & $(M^{-1}\cdot 10^7)$ & $(\cdot 10^{-13})$ & $(M^{-1}\cdot 10^{-5})$\\
         \hline
         ACTTCT & $6$ & $11.0(1.3)$ & $12.3(21.3)$& $14.0$\\
         ACTT & $6$ & $0.3(0.1)$ & $6.0(1.0)$ & $0.2$\\
         ACT & $6$ & $13.0(7.0)$ & $0.3(0.6)$ & $0.4$ \\
    \end{tabular}
    \caption{Table with the bindings constants obtained from a fit with Eq.~\ref{eq:theta_multi} to $\Theta$ in Fig.~\ref{fig:superselectivity_noinert}b yields the fitting parameters $K_{\rm A}/K_{\rm intra}$ and $K_{\rm intra}$. The numbers in brackets indicate the fitting error. A division of the two fitting parameters yields $K_{\rm A}$.}
    \label{tab:Fitting_par_Fig_3}
\end{table}

We test these theoretical predictions by measuring $\Theta$ for DNA nanostars with valency $k=6$ and interaction strength of the individual arms of $\Delta G^0 = -5~\rm{k_BT}$ and compare the results to a DNA nanostar with equal valency $k=6$ but stronger interaction strength $\Delta G^0 = -7~\rm{k_BT}$. The data of $\Theta$ as a function of $\sigma_{\rm R}$ for these DNA nanostars with two different sticky ends is presented in Fig.~\ref{fig:6bp_vs_4bp}. 
Comparing the two results, we immediately notice that $\Theta_{\rm ACTT}$ shifts to higher $\sigma_{\rm R}$ compared to $\Theta_{\rm ACTTCT}$, in line with the theoretical predictions and with experiments performed on colloidal systems \cite{Linne2021}.

To investigate if and how the selectivity and binding constants vary between these two DNA nanostars, we fit Eq.~\ref{eq:theta_multi} with $K_{\rm A}$ and $K_{\rm intra}$ as fitting parameters as described in the previous section. We note that only the first four data points of the $4~$bp sticky end were used in the fit to capture the non-linear transition as accurately as possible, because it determines the maximum selectivity of the system.
We find that weakening the interaction strength indeed makes the DNA nanostars more superselective. This means that, since DNA nanostars with an interaction strength of zero will not bind there is an optimal interaction strength to achieve highest superselectivity. Comparing the equilibrium constants $K_{\rm A}$ and $K_{\rm intra}$ of $k=6$ for the $4~$bp and $6~$bp sticky ends (reported in Fig.~\ref{fig:6bp_vs_4bp}b), we notice that whereas the equilibrium constant for binding the first arm, $K_\mathrm{A}$, shows a difference of one order of magnitude, the values for the equilibrium constant associated to binding subsequent arms, $K_\mathrm{intra}$, are similar for the two different sticky end lengths. This is a puzzling observation, because one would expect $K_\mathrm{A}$ and $K_\mathrm{intra}$ to scale similarly with the binding strength, $K\propto \exp (-\beta f)$, with $f$ the binding free energy and $beta$ as 1/$k_\mathrm{b}T$ and indicates that the simplest version of the model is either incomplete, misses relevant interactions, or is inconsistent.

\section*{Theory about optimum in valency and binding strength}

\begin{figure}
    \centering
    \includegraphics[width=\linewidth]{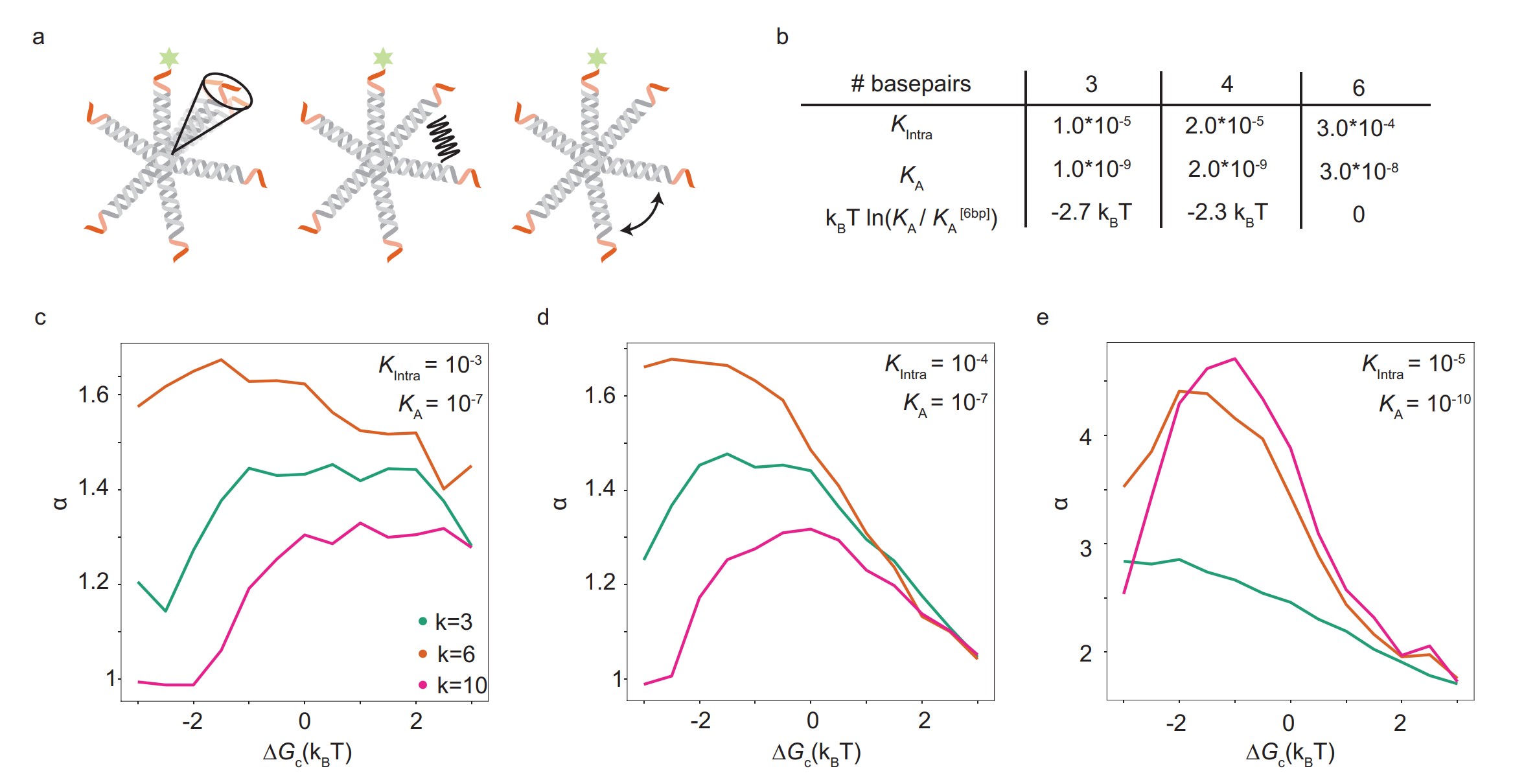}
    \caption{\textbf{Model expansion} a) Soft interactions between the ligands may significantly affect the binding rates. These interactions may have an entropic or energetic origin. b) Experimental fit values for the rate constants $K_\mathrm{A}$ and $K_\mathrm{intra}$. The ratio $K_\mathrm{A}/K_\mathrm{intra}$ was taken to be constant, because they are expected to be equally affected by the binding strength. The difference in binding free energy is given in the third row, compared to the 6 bp sticky end. c) Simulation results show the maximal achievable selectivity as a function of pair-interaction strength $\Delta G_\mathrm{c}$, which is `cooperative' as $\Delta G_\mathrm{c}<0$ and `competitive' as $\Delta G_\mathrm{c}>0$. The on rates are highest in the left figure (which corresponds best to the experimental data), and lowest in the right figure. There is an optimal $\Delta G_\mathrm{c}$ corresponding to a weak cooperative interaction for low rates, and a weak competitive interaction for higher rates.}
    \label{fig:Fig4}
\end{figure}

The data demonstrate that a maximal selectivity is achieved for $k=3$ arms, and suggest that $K_\mathrm{intra}$ is dependent on valency, but not on the binding strength.To understand these puzzling observations it is helpful to consider the average number of bound arms of a bound DNA nanostar,
$$
    \langle n \rangle = 1 + \frac{k - 1}{1 + (\sigma_{\rm R} A  K_{\rm intra})^{-1}}.
    \label{eq:arms}
$$
where n is the number of bound arms, which is effectively 1 if few receptors are available, and reaches the asymptote of $k$ in the high $\sigma_{\rm{R}}$-limit.
The derivation of this expression follows the same assumption as was used in Eq. \ref{eq:theta_multi} (that the arms act independently), except that one only has to consider the microstates of a single DNA nanostar. The receptor concentration where half of the arms are bound on average depends only on $K_{\rm{intra}}$ and not on the concentration of DNA nanostars. As long as $\langle n \rangle \approx 1$, the DNA nanostars behave effectively as monovalent particles, and the selectivity $\alpha \approx 1$. To a good approximation $\alpha \approx \langle n \rangle$, up to a point where $\alpha$ starts to drop to zero, because of saturation effects (of the surface, or depletion of the bulk), or to 1, if all available receptors become occupied. Whether a system behaves superselective or not, depends sensitively on this crossover density $\sigma_{\rm{R}}^*$, with $\sigma_{\rm{R}}^*A = K_{\rm{intra}}^{-1}$. From this simple model, one can conclude that there are four distinct regimes, namely 1) for small $\sigma_{\rm{R}}\ll \sigma_{\rm{R}}^*$: particles bind with 1 arm on average and behave effectively monovalent, so $\alpha=1$, regime 2) around $\sigma_{\rm{R}} \approx \sigma_{\rm{R}}^*$: the system becomes superselective, with $\alpha \approx \langle n \rangle$, regime 3) for higher $\sigma_{\rm{R}} > \sigma_{\rm{R}}^*$: $\alpha$ drops to 1 when all the available receptors become occupied, and regime 4) for high $\sigma_{\rm{R}} \gg \sigma_{\rm{R}}^*$: $\alpha$ drops to zero due to saturation effects, which happens either when the bulk becomes depleted, or the surface becomes too crowded. Regime 1 and 4 should in principle always be achievable for any type of particle, but the interesting regime 2) is only visible if saturation effects are not important for $\sigma_{\rm{R}} < \sigma_{\rm{R}}^*$. This explains why a large number of arms $k$ can be disadvantageous, because the larger $k$ is, the sooner particles bind, and the sooner saturation effects appear. Only if one can successfully prevent early binding via other principles (steric repulsions, entropy barriers) could one achieve the maximum selectivity of $\alpha = k$.

The analytical expressions contain in principle four tunable parameters, the two rate constants $K_\mathrm{A}$ and $K_\mathrm{intra}$, the number of ligands $k$, and a maximum number of DNA nanostars per area $N_\mathrm{max}$ that determines the normalisation of $\Theta$. The value of $N_\mathrm{max}$ determines when saturation effects become important, and is dependent on the size of the DNA nanostars, setting the maximum packing fraction. In scenarios with very low bulk concentrations, $N_\mathrm{max}$ is the available number of DNA nanostars in solution. The rate constants $K_\mathrm{intra}$ and $K_\mathrm{A}$ are dependent on the binding strength of the ligands and their configurational degrees of freedom. Comparing now the binding probability of three different types of particles, with 3 ligands, 6 ligands and 10 ligands, having identical ligands but only differing in the number of arms, we would expect all parameters to be identical, except for the number of arms $k$. However, we observe (as described above) that the experimental data cannot be fitted with a single set of parameters, and only show reasonable agreement if we fit a different $K_\mathrm{intra}$-value to each set. This observation challenges the interpretation of $K_\mathrm{intra}$, and seems to be a clear indication that there are effects not captured by the model we used so far. Therefore we make a minimal extension of the model to estimate the potential influence of self-interactions and pair-interactions between the ligands.

Soft pair interactions between the ligands and self-interactions could alter the binding kinetics, for example via steric interactions\cite{watzlawek1999phase}, ion-bridging, electrostatic interactions\cite{raspaud1998precipitation,sing2013effect}, or stiffness of the arms and joints\cite{xing2018microrheology,stoev2020role} (see Fig. \ref{fig:Fig4}a). The strength of these effects would depend on the total number of ligands $k$, being more important for particles with many ligands. Keeping the microscopic origin of these effects unspecified, we include a mesoscopic parameter $\Delta G_\mathrm{c}$ to represent these effects in the form of a Gibbs free energy, calling the effect cooperative if $\Delta G_\mathrm{c}<0$, competitive if $\Delta G_\mathrm{c} > 0$, and $\Delta G_\mathrm{c}=0$ representing the original analytical model where the ligands act independently. This parameter alters the transition rates between the states in the following way:
\begin{eqnarray}
    \frac{c_{12}}{c_{21}} &=& \frac{k-1}{2} K_\mathrm{intra} e^{- \Delta G_\mathrm{c}} \nonumber\\   
    \frac{c_{23}}{c_{32}} &=& \frac{k-2}{3} K_\mathrm{intra} e^{- 2\Delta G_\mathrm{c}} \nonumber\\ 
    &\vdots& \nonumber\\
    \frac{c_{n,n+1}}{c_{n+1,n}} &=& \frac{k-n}{n+1} K_\mathrm{intra} e^{-n\Delta G_\mathrm{c}}\nonumber 
\end{eqnarray}
with $c_{nm}$ the rate constant that a DNA nanostar makes a transition from a state with $n$ arms bound to a state with $m$ arms bound, with $m = n \pm 1$. The combinatorial factor is the number of available ligands in state $n$ able to bind, divided by the number of arms that are bound in state $n+1$ and able to detach. The correction of the rate constant increases with $n$, representative of an attractive pair interaction, where $n$ bound arms each interact with $n-1$ other bound arms.

\begin{table}[h]
    \centering
    \begin{tabular}{|c||c|c|c|}
         \hline
         \# ligands & 3 & 6 & 10 \\
         \hline
         $\Delta G_\mathrm{c}$ ($k_\mathrm{B}T$) & -0.5 & 0 & 1.0 \\
         \hline
         %$K_\mathrm{intra}$ & $3\cdot10^{-4}$ &  &  \\
         %\hline
         %$K_\mathrm{A}$ & 3\cdot10^{-8}$ &  &  \\
         %\hline
    \end{tabular}
    \caption{Parameter values that fit the model to the experimental profiles. All three types of particles were fitted with $K_\mathrm{intra} \approx 3\cdot10^{-4}$ and $K_\mathrm{A} \approx 3\cdot10^{-8}$.}
    \label{tab:Fit_values_Delta_G_c}
\end{table}

Now we can fit this model to our experimental data to extract single values for $K_\mathrm{A}$ and $K_\mathrm{intra}$ from the experimental data, with a small difference in $\Delta G_\mathrm{c}$, given in Tab./ \ref{tab:Fit_values_Delta_G_c}. 
 The relative differences in $\Delta G_\mathrm{c}$ indicate that the ligands of 3-armed DNA nanostars experience less competition than 10-armed DNA nanostars, with a difference of about $1.5\ k_\mathrm{B}T$ in the effective pair-interaction between the ligands.

6-armed DNA Nanostars with 3 and 4 basepair sticky ends have a lower binding probability than those with 6 basepair sticky ends, as shown in Fig. \ref{fig:6bp_vs_4bp} The length of the sticky end is expected to affect the unbinding rates \cite{Martinez-Veracoechea2011} $k_\mathrm{off}\propto \exp(\beta f)$, with $f$ the binding free energy of one ligand. Therefore the length of the sticky end is expected to influence $K_\mathrm{intra}$ and $K_\mathrm{A}$ by the same factor, as they both depend on $1/k_\mathrm{off}$. Assuming that the pair interactions are not influenced by the sticky end (or negligibly), $\Delta G_\mathrm{c}$ should be the same for these particles. Curve fits are found for the parameters given in the Table of Fig. \ref{fig:Fig4}b. The relative differences in binding free energy between the different ligands is estimated by $k_\mathrm{B}T\ln(K_\mathrm{A}/K_\mathrm{A}^{[6\mathrm{bp}]})$.

%\begin{table}[h]
%    \centering
%    \begin{tabular}{|c||c|c|c|}
%         \hline
%         \# basepairs & 3 & 4 & 6 \\
%         \hline
%         $K_\mathrm{intra}$ & $1.0\cdot10^{-5}$ & $2.0\cdot10^{-5}$ & $3.0\cdot10^{-4}$ \\
%         \hline
%         $K_\mathrm{A}$ & $1.0\cdot10^{-9}$ & $2.0\cdot10^{-9}$ & $3.0\cdot10^{-8}$ \\
%         \hline
%         $k_\mathrm{B}T\ln(K_\mathrm{A}/K_\mathrm{A}^{[6\mathrm{bp}]})$ & $-2.7\ k_\mathrm{B}T$ & $-2.3\ k_\mathrm{B}T$ & $0$ \\
%         \hline
%    \end{tabular}
%    \caption{Parameter values for 6-armed nanostars having 3, 4, and 6 basepair sticky ends. All three types of particles were fitted with $\Delta G_\mathrm{c}=0$.}
%    \label{tab:fit_values_K_intra_K_A}
%\end{table}

The parameter values in the Table of Fig. \ref{fig:Fig4}b suggest that a lower $\Delta G_\mathrm{c}$, meaning more cooperation between the ligands, is favorable. However, a numerical exploration of parameter space seems to suggest that there is an optimal value of $\Delta G_\mathrm{c}$, and that there are situations where competition between the ligands results in a higher selectivity than cooperation. Figs. \ref{fig:Fig4}c, \ref{fig:Fig4}d, and \ref{fig:Fig4}e show the maximal selectivity of a DNA nanostar as a function of $\Delta G_\mathrm{c}$ for different values of the binding rates, being highest in the left figure and lowest in the right figure. Intuitively one expects cooperation to be favorable to achieve a higher selectivity, because it could make the average number of bound arms $\langle n \rangle$ depend more sensitively on the receptor density $\sigma_\mathrm{R}$ (Eq. \ref{eq:arms}), potentially even leading to a first order phase transition, where $\langle n \rangle$ makes a sudden jump at a critical value of $\sigma_\mathrm{R}$. Although this transition would drastically change the binding dynamics from solution, the availability of receptors would also drastically decrease, as the particles already bound would rapidly reduce the number of available receptors, and prevent particles from solution to bind. The effect of cooperation on the binding probability $\Theta$ is not as obvious as the effect on $\langle n \rangle$, and the simple relation between $\alpha$ and $\langle n \rangle$ that we found earlier does no longer hold in this scenario. For the parameter values that match the experimental data best, the results show that a small competition is actually favorable over a cooperative interaction, and that the maximal selectivity depends more sensitively on $\Delta G_\mathrm{c}$ for DNA nanostars with more arms. The trivalent DNA nanostars have the highest selectivity and are least affected by the value of $\Delta G_\mathrm{c}$. This observation suggests two important biological reasons why e.g. viruses would use a small number of ligands to select their host. (1) In a crowded environment, a smaller number of arms may result in a higher selectivity, and (2) the selectivity is stable under changing environmental conditions that influence the interactions between the arms.

\section*{Discussion}

Specific data would be required to interrogate the microscopic origin of the pair-interactions, and fix the zero-point of the Gibbs free energy $\Delta G_\mathrm{c}$. As long as the DNA nanostars do not bind with more arms than 2, such that the selectivity $\alpha\leq2$, the Gibbs free energy $\Delta G_\mathrm{c}$ would simply rescale $K_\mathrm{intra}$, as is the case for our experimental results. In this case, there is effectively only one rate constant between the first two bound states, which allows freedom where to set $\Delta G_\mathrm{c}=0$ because there is only one transition with two variables, $K_\mathrm{intra}$ and $\Delta G_\mathrm{c}$. 
Only in the parameter regime where $\alpha > 2$ could one potentially validate the specific adaptation of the transition rates, and gain more information about the type of mechanism responsible for $\Delta G_\mathrm{c}$. It is to be expected that the transition rates will depend on the number of bound arms, not only because of combinatorial reasons, but also because of the specific energy of the configuration, depending on the type of interaction. Data on the binding probability of particles with a selectivity $\alpha > 2$ could shed light on the microscopic origin and strength of the pair-interactions.

In summary, we have interrogated the effect of valency and binding strength on the selectivity of multivalent objects, with a limited valency $k= 3, 6, 10$, and found that both the valency and binding strength have an optimal value to achieve maximal selectivity. We observed that DNA nanostars with 3 ligands can be more selective than those with 6 and 10 ligands, and can explain this from the fact that particles with more ligands have a larger binding rate from solution, such that the surface saturates sooner, hindering binding. After comparing the observations to the theoretical model, we also concluded that there may be relevant pair-interactions between the ligands. Including this effect at a mesoscopic level, we found agreement with our observations. By exploring parameter space with simulations, we found that the selectivity has a maximal value for an optimal strength of the cooperative interactions, and that there are even conditions where weak competitive interactions are optimal. These conditions include the parameters that were found from fits with the experimental data. 

Based on our results we can formulate several design rules for maximizing selectivity under different experimental conditions: Aligned with earlier conclusions\cite{Martinez-Veracoechea2011}, a maximal selectivity can be found at infinite dilution of the bulk solution, where $\alpha \rightarrow k$. In this limit, a larger valency always leads to a larger selectivity. However, this limit may not be  experimentally accessible, as the equilibration time also drastically increases, and may not be relevant in a biological context, where concentrations are finite and selectivity needs to be established within a certain time interval. At finite concentrations, there is an optimal value of the valency and binding strength, which can be estimated with existing theoretical models, once the binding rates are measured. Weak interactions between the ligands complicate the picture, and have a larger influence the larger the valency is. Surprisingly, we find cooperative effects to be unfavorable for the experimental conditions, and only favorable if the on-rate from solution is sufficiently low, requiring a larger entropy barrier (by diluting the bulk solution) or stronger energy barrier (by e.g. steric hindrance) between the free and bound state. In conclusion, increasing the valency of a particle may actually lower the selectivity, and make the particles more sensitive to unwanted pair-interactions between the ligands. A limited valency may be favorable for maximizing the selectivity and robustness to environmental changes that affect the ligand-ligand interactions.

\section*{Materials and Methods} {\label{sec:methods}
%\begin{figure}[htbp]
%    \centering
%    \includegraphics[width=\textwidth]{Figures/Methods/20211031_Fig1_2.eps}
%    \caption{\textbf{DNA nanostar design} a) Four ssDNA with a sticky end form the star shaped structure during hybridization. Four free $A$ bp at the centre of the nanostar enable full flexibility of the arms. One sticky end possesses an ATTO488 dye for fluorescent imaging. b) DNA electrophoresis allows for the visualization of the final DNA star products after hybridization. The fluorescent bands shift upwards and confirm the formation of larger DNA nanostructures. Each lane contains faint lower bands that represent incompletely hybridized side products.}
%    \label{fig:nanostar_design}
%\end{figure}
\paragraph{DNA nanostar hybridisation}
All DNA strands were purchased from Integrated DNA Technologies Inc (IDT), resuspended in Tris buffer (pH8) and stored at $-20^{\circ}C$. To achieve for example tetravalent DNA nanostars with four sticky ends and one fluorophore, we mixed the four DNA strands $X1$, $X2$, $X4$ and $X5$ in equal molar ratios and annealed the mixture to $95^{\circ}C$ for $10~$min and then cool it down at a rate of $0.2^{\circ}C/$min to $4^{\circ}C$, see Supplemental Table 1 and see Supplemental Fig. 1a. The annealing took place in a Thermocycler and a final concentration of $0.5~\mu$M. The final product was stored at $4^{\circ}C$. For the experiments, we diluted the desired concentration of DNA nanostars and receptors in Tris Acetate-EDTA-NaCl (TAE,$100~$mM NaCl, pH=$8$) and $10~$mM magnesium chloride (MgCl). \\
To verify the hybridization of the DNA nanostars, we performed DNA electrophoresis. The sample consisted of $10~\mu$L of $0.5~\mu$M DNA nanostars and we loaded the sample on a $1$\% agarose gel. After $30~$min at $100~$V we took an image of the gel, see  supplemental figure 1b. The fluorescent bands correspond to DNA nanostars: the higher the band, the larger the DNA nanostar nanostructure. Lower bands result from incomplete hybridizations. The intensity of the upper bands is significantly higher, confirming the successful formations of the DNA nanostars.

\paragraph{DNA functionalised supported lipid bilayer}
We studied the DNA nanostar adsorption in solution in a flow channel. The supported lipid bilayer (SLB) consisted of 18:1 $1,2$-dioleoyl-sn-glycero-$3$-phosphocholine (DOPC, [Avanti Polar lipids], stored in chloroform). To obtain the SLB, we first made small unilamellar vesicles (SUVs) from DOPC lipids. To do so, we added the desired volume of lipid to a glass vial and let it dry overnight in a vacuum desiccator. Subsequently, we resuspended the lipids in TAE-NaCl buffer and extruded the solution with an Avanti mini extruder through a membrane with pore size of $30~$nm (Avanti Polar lipids). The microscopy slides and coverslips were sonicated at $30~$min each in $2$\% Hellmanex solution, acetone ($>99.9$\%) and potassium hydroxide solution (KOH,$1~$M,[Merck]). Between each change of chemical we rinsed the glass ware [VWR] with milliQ water. Before use, the slides and coverglasses were blown dry with nitrogen. Parafilm stripes confined the flow channel and glued the microscopy slide and a coverslip together. Subsequent annealing at $125^{\circ}C$ let the Parafilm melt and bound the microscopy slide and coverslip together yielding $(1 \times 22~)$mm rectangular flow channels. To obtain SLBs in the flow channel, we injected SUVs and after $30~$ min at room temperature, we washed out the excess SUVs with buffer and added DNA of the desired concentration.  
\paragraph{Data acquisition and analysis}
To image the DNA nanostar adsorption on the target surface we used Total Internal Reflection Microscopy (TIRF) on an inverted fluorescence microscope (Nikon Ti2-E) upgraded with an azimuthal TIRF/FRAP illumination module (Gata systems,iLAS 2) equipped with a $100\times$ oil immersion objective (Nikon Apo TIRF, $1.49$NA). Each DNA nanostar possesses an Atto488 dye and each receptor features a Cy3 dye. Therefore, we used laser excitations with wavelength $488~$nm and $561$nm and detect the emitted fluorescent signal (EM-CCD Andor iXON Ultra 897). For each binding probability we measure for $7$ different $\sigma_{\rm R}$ the intensity of the DNA nanostars $I$ to obtain the full range of adsorption from unbound to bound. A negative control with $\sigma_{\rm R} = 0~\mu \rm m^{-2}$ defines the background signal $I_{\rm back}$. The maximum intensity $I_{\rm max}$ provides a reference for normalization. The monovalent DNA nanostars were not measured until saturation due to practical constraints. Therefore, we normalized the monovalent signal with the maximum signal of $k=6$ of the same sticky end. After the acquisition of the DNA nanostar adsorption in equilibrium, the acquired signal is corrected and normalized yielding the binding probability:
$$
    \Theta = \frac{I - I_{\rm back}}{I_{\rm max} - I_{\rm back}}.
$$
For the image processing we used a combination of ImageJ and python. 

\paragraph{Simulation method} The system is described as a reaction network, consisting of transitions between different bound states, and the free state in solution, with corresponding rate constants. This network is stochastically evolved using a kinetic Monte Carlo algorithm, according to Gillespie \cite{gillespie1977exact}. After equilibration, the coverage is obtained as the average over a large number of iterations ($n \gg 10^4$).

%FRAP measurements revealed the binding kinetics of the nanostar adsorption. The acquired data was normalised with respect to the initial intensity before bleaching. Furthermore, we corrected for the bleaching of the fluorophores over time. 

}

%\showmatmethods{} % Display the Materials and Methods section

\section*{Acknowledgements} 
We thank Ramon van der Valk for technical help, Jérémie Capoulade for help with microscopy and Constant Tellinga for his pioneering theory work on surface binding by nanostars. C.L., L.L., and D.J.K. acknowledge support from the Netherlands Organization for Scientific Research (NWO/OCW), as part of the Gravitation Program: Frontiers of Nanoscience.

\bibliographystyle{unsrtnat}
\bibliography{references}  %%% Uncomment this line and comment out the ``thebibliography'' section below to use the external .bib file (using bibtex) .





% \end{thebibliography}

\end{document}